\documentclass[manuscript]{aa}
\usepackage{graphics}
\thesaurus{05(08.01.1;08.12.1;08.09.3;10.15.1 Coma Berenices)}
\begin{document}
\title {Lithium in the Coma Berenices Open Cluster}

\author{A. Ford
\inst{1}
\and R.D. Jeffries 
\inst{1}
\and D.J. James
\inst {2}
\and J.R. Barnes 
\inst {2}}
\offprints {A. Ford}
\institute{Department of Physics, Keele University, Keele, Staffordshire, ST5 5BG. UK \\
email: af@astro.keele.ac.uk or rdj@astro.keele.ac.uk
\and
School of Physics and Astronomy, University of St. Andrews, North Haugh, St Andrews, Fife, KY16 9SS. UK \\
email: djj@st-andrews.ac.uk}
\date{Received; accepted}
\authorrunning{A. Ford et al.}
\maketitle
\begin {abstract}

Lithium abundances, radial velocities, and rotational velocities are 
reported for 20 candidate F, G and K stars of the sparse Coma Berenices open
cluster. All the stars are proper-motion selected, and our radial
velocities support the membership credentials of at least 12 of the
candidates. Combining our data with that in the literature, we have a
close-to-complete census of Li abundances for late-type stars with
$M_{V}<5.8$. These data show that the Li-depletion pattern in Coma Ber
is similar, but not identical, to that in the Hyades cluster, which has
a similar age but higher metallicity.  Several Coma Ber F stars have
suffered significantly more Li depletion than their counterparts in the
Hyades. The G and early-K stars of Coma Ber have undergone less Li
depletion than those of the Hyades, but much more than that predicted by
standard evolutionary models featuring only convective mixing.
This provides strong evidence for additional mixing and Li depletion 
operating in these stars during their first 400\,Myr on the main
sequence, amounting to 0.3 dex at 6000\,K and rising to 0.8-1.2 dex at 
5400\,K. We find that 4 of the radial-velocity non-members are among
a small number of low-mass stars which were previously reported as part of an
extra-tidal moving group associated with the Coma Ber cluster. As a
result, we now find that the luminosity function of this moving group is
indistinguishable from that of the central cluster. It is
uncertain whether there may be a significant number of stars with
even lower masses among the moving group.  

\end {abstract}


\section {INTRODUCTION}

Lithium is the only metal which was formed in significant quantities in the
big bang and yet can easily be burned in stellar interiors. Knowledge
of the primordial lithium abundance could, in principle, be used
to constrain big-bang nucleosynthesis models and the universal baryon
density (Yang et al. \cite{yang84}). Li abundances in Population II
stars might represent this primordial abundance (Bonifacio \& Molaro
\cite{bonifacio97}), but it is debatable whether stellar mixing and
processing have depleted lithium from a higher initial value
(Deliyannis \& Ryan \cite{deliyannis97}; Ryan et al. \cite{ryan99}).  Standard stellar-evolution models, incorporating only
convective mixing, predict little Li depletion in Population II
stars. However, non-standard processes such as microscopic diffusion,
rotation-induced turbulence, or gravity waves might have reduced the
surface Li by factors of 2-3 (Pinsonneault et al. \cite{pinsonneault99}).

Much of the support for non-standard processes is established in
observations of young clusters, where rapid rotation and
loss of angular momentum may well drive additional mixing.  
The standard models cannot explain the ``Boesgaard Gap'' of
Li-depleted mid-F stars in clusters such as the Hyades (Michaud 
\cite{michaud86}; Charbonneau \& Michaud \cite{charbonneau88}), the
spread in lithium abundances in younger clusters such as the Pleiades
(Soderblom et al. \cite{s93a}) and $\alpha$ Per (Balachandran et
al. \cite{balachandran96}), or the evidence that Li depletion
continues in main-sequence stars, even though the bases of their
convection zones (CZs) are too cool to burn Li (Jeffries \& James \cite{jj99}).

Late-type stars in the Coma Berenices open cluster offer an important
test of Li depletion models. With an age of 400-500\,Myr (Philip et
al. \cite{philip77}; Lyng\aa\ \cite{lynga87}) Coma Ber is similar to the well
studied Hyades and Praesepe clusters (ages
$\simeq$600--700\,Myr -- Mermilliod 1981). Importantly, these open clusters have a set of
consistently-determined spectroscopic iron abundances: Coma Ber
[Fe/H]=$-0.052\pm0.026$; Hyades [Fe/H]=$+0.127\pm0.022$; Praesepe
[Fe/H]=$+0.038\pm0.039$ (Boesgaard \& Friel \cite{boesgaard90}; Friel
\& Boesgaard \cite{friel92}). Because the metallicity in Coma Ber is
lower than that of the Hyades, its G and K stars should have shallower
CZs at similar effective temperatures ($T_\mathrm{eff}$), and are
predicted to undergo {\em significantly less} Li depletion during the
pre-main-sequence (PMS) convective-mixing phase. As CZ-base
temperatures are too cool to burn Li in main-sequence stars {\em
hotter} than $\simeq$5000\,K, continuing Li depletion in these
stars is not predicted by standard models. On this basis we would
expect the Li-depletion patterns in these three post-ZAMS clusters to
be very different, strongly dependent on metallicity, and set during
their PMS evolution. Furthermore, because the Pleiades, with an age of
100Myr, has a consistently-determined metallicity
([Fe/H]=-0.034+/-0.024), we can compare it with the older Coma Ber
cluster to study the presence, timescales and mass dependence of any
non-standard mixing and Li-depletion mechanisms for F, G and K stars
which have reached the ZAMS. 

Jeffries (1999 - hereafter \cite{j99}) obtained Li abundances for a
small sample of F and G stars in the Coma Ber cluster. It was found
that the depletion pattern was very similar to that of the Hyades,
with only a hint of higher Li abundances in stars cooler than
5700\,K. \cite{j99} concluded that the Li depletion in a range of
clusters (including Blanco 1 and the Pleiades, as well as the Hyades
and Coma Ber) is ordered according to the cluster age, rather than
metallicity. This is contrary to the predictions of standard models. 
However, non-standard mixing is {\em in addition to} PMS convective
mixing, so clear differences between Coma Ber and the
Hyades should remain. It is, therefore, still difficult to explain the
Coma Ber Li abundances unless convective PMS Li depletion is strongly
inhibited in all of the sample clusters. The conclusions of \cite{j99}
were tempered by the small sample size -- 10 single members, 4
spectroscopic binaries and one possible non-member -- and the
relatively low resolution (0.5\AA) spectra used to study blended Li lines.

Increasing the sample size is difficult because the cluster is spread
over a wide area on the sky, is sparse, and has few known low-mass
members. A new proper-motion study has recently been published by
Odenkirchen et al. (\cite{odenkirchen98}, hereafter O98), who have identified several
low-mass candidates which were not observed in \cite{j99}. In addition,
they find a core-halo spatial distribution and a co-moving group of
extra-tidal stars. There is some evidence (albeit based on very small
number statistics) that the mass function is still increasing towards
lower masses in those stars at large distances from the cluster centre.
Towards the cluster core the mass function appears sharply
truncated just below 1M$_{\odot}$. These observations would agree with
the idea that low-mass stars undergo mass segregation and
preferential evaporation from young clusters, would explain the paucity
of low-mass stars in previous surveys, and would predict a large number
of stars with still lower masses at $>$5$^{\circ}$ from the
cluster centre.

In this paper we determine radial velocities for Coma Ber
cluster-member candidates proposed in O98
as well as a few other candidates from earlier proper-motion surveys
(Trumpler \cite{trumpler38}). We use these results to filter the
candidate members, then determine Li abundances for all the objects
and re-examine the conclusions of \cite{j99} using an enlarged sample of stars.

\section {OBSERVATIONS AND ANALYSIS}
\subsection {Target Selection}

The main source of new targets in this work (compared with \cite{j99})
is the proper-motion survey of O98, based upon Hipparcos (Perryman et
al. \cite{perryman97}) and ACT (Urban et al. \cite{urban98})
proper-motion measurements within a 20$^{\circ}$ radius of the nominal
Coma Ber open cluster centre.  For the brighter stars ($V<8.5$) in the
Hipparcos catalogue, proper motions alone were used to select members,
whereas fainter stars ($V<10.5$) were chosen from the ACT catalogue
using proper motions in
conjunction with photometric constraints. A total of 34 stars with
$V<10.5$ were identified in this way (20 of them had $V<8.5$). This
catalogue is expected to be reasonably complete, although a
cross-check with the number of cluster members predicted by the
excess stellar density over background counts suggests that perhaps as
many as 40 per cent of members could have been missed. The reasons for
this are two-fold: first, the instantaneous proper-motion
measurements of binary systems may have given misleading information
about their membership. This incompleteness is probably magnitude-independent.
 Second, for fainter targets ($V>8.5$), the Tycho photometry used
for photometric selection can be quite uncertain. This, together with
the severe photometric constraints used by O98 could
have led to additional incompleteness among the faintest objects. 

For the main purpose of this paper incompleteness is not a great
problem, except that it limits the number of objects for which we can
measure Li abundance. The targets observed here were selected as
proper-motion members of the cluster by O98 and all are F, G or K type, with
9.10$<V<$10.78. Including those stars observed by \cite{j99}, we have
studied all the stars from O98 which are mid-F type or
later. Five faint proper-motion candidates (Ta 3, Ta 13, Ta 19, Ta 20
and Ta 21) were also selected from Trumpler (1938), four of which (the
exception being Ta 20) have been reported to show photometric modulation 
which is consistent with their identification as youthful cool cluster members by Marilli et al. (\cite{marilli97}). 
These targets were faint enough ($10.36<V<11.18$) that their absence from
the catalogue of O98 does not argue strongly against their
membership. A list of the new targets observed in this paper is given
in Table~1.

\subsection {Spectroscopy}
 
The new spectroscopic data acquired for this paper 
were obtained at the 2.5m Isaac Newton Telescope (INT) on 1998
November 26, and at the 4.2m William Herschel Telescope (WHT) on 1998
November 28--29, 1998 December 06--08, and 1999 December 22--23 inclusive.

The INT data were taken with the same instrumental setup as described
by \cite{j99} and analysed in exactly the same way to derive
heliocentric radial velocities and the equivalent widths (EWs) of the
\ion{Li}{i} 6707.8{\AA}/\ion{Fe}{i} 6707.4{\AA} blend. The instrumental
resolution was 0.5\AA\, and the data consisted of a pair of spectra for
each target which were co-added to give a signal-to-noise ratio (s/n) of
approximately 200 per 0.22\AA\ pixel.

The WHT data were taken with the Utrecht Echelle Spectrograph (UES) in
cross-dispersing mode at the Nasmyth focus of the telescope.  The
detector used was a SITe1 2048x2048 CCD.  The UES was set up with a 31
grooves/mm grating and a central wavelength of 5875{\AA}, so that a
wavelength range from 4600{\AA} to 9000{\AA} was covered at a dispersion of
0.066{\AA} per 24$\mu$m pixel in the region of the \ion{Li}{i} 6708{\AA}
line. The resolving power was 45000 for the 1 arcsecond slit width we
used.

\begin{figure}
\resizebox{\hsize}{!}{\includegraphics{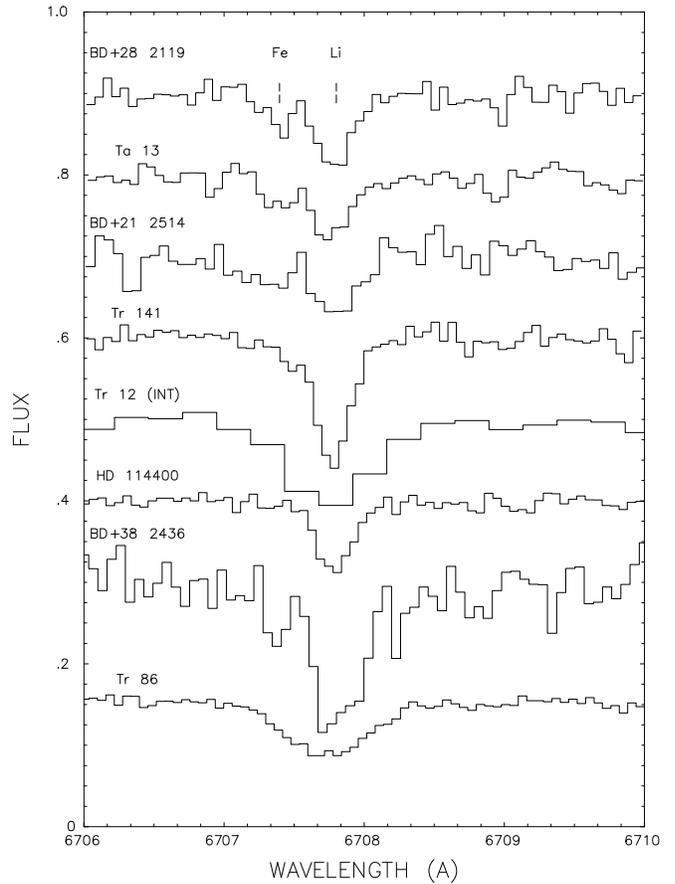}}
\caption{Spectra in the region of the 6708{\AA} line for a selection of
our sample stars. The spectra have been normalised to unity and offset for
clarity. Spectra have been corrected for their heliocentric radial velocities.}
\label{spec}
\end{figure}

The usual calibration frames, tungsten flat-fields and B stars were
taken, as well as Thorium-Argon lamp exposures. The data were reduced
using the Starlink {\sc echomop} and {\sc figaro} packages,
which included bias-subtraction, flat-fielding, subtraction of
scattered light and wavelength calibration. The s/n of these spectra
in the radial-velocity orders (see below) and around the \ion{Li}{i}
6708{\AA} line were in the range 60 to 170 per extracted spectrum
pixel.  Radial velocities (RVs) were measured using cross-correlation
techniques in the spectral range 5490 to 5584{\AA} and 5940 to
6150{\AA}, which contained many sharp, neutral metal lines and very
little telluric contamination. The
standard templates used to determine heliocentric RVs were HD 222368,
HD 32963, HD 126053 and HD 693. Spectra for several of our objects showing
the region around the 6708{\AA} Li line are shown in Fig.\,\ref{spec}.

The RVs obtained from the INT and WHT spectra are detailed in Table 1.
The errors in the INT velocities, taken with a Cassegrain spectrograph,
are dominated by wavelength stability. Repeat exposures and
cross-correlation of standards taken on different nights yield an
estimated internal error of $\pm 2$\,km\,s$^{-1}$, which is larger than
the external error of putting the velocities onto the standard system.
UES is mounted on a Nasmyth platform and has much higher
resolution. The same experiments on standard stars observed at
different times yield internal error estimates of less than
0.3\,km\,s$^{-1}$, with a likely external error of about
$\pm1$\,km\,s$^{-1}$. We quote the approximate sum of internal and external errors
in Table~1, as this is appropriate for comparing with other data
(in particular that from J99), although this might slightly
over-estimate the internal scatter within the WHT data.

Projected equatorial velocities were estimated for all the stars by
cross-correlating the same spectral orders used for RV determination with those
of slowly-rotating, chromospherically-inactive stars of a similar
spectral type. The width of the cross-correlation peak was calibrated
against $v_{e}\sin i$ by artificially broadening the template-star
spectra with a limb-darkened rotation profile (Gray 1992) and cross-correlating
against other template stars. A limb-darkening coefficient of 0.6 was used. For the INT spectra only an upper limit
of 15\,km\,s$^{-1}$ could be determined. For the WHT data, with better
resolution, we could measure $v_{e}\sin i$ down to a limit of
6\,km\,s$^{-1}$ and several stars showed broadening greater than
this. From numerical experiments we estimate accuracies of order 10\%
for these values, which are listed in Table 1.

\subsection {Lithium Abundances}

The main purpose of our paper is to compare the Li abundances of cool stars in the Coma Ber open cluster with those in other clusters such as the Pleiades and Hyades. To do this is {\em absolutely essential} that consistent abundance determination techniques are used for all the stars involved in the comparison, including atmospheric models, deblending techniques and temperature scales (see Balachandran et al. 1995). Because all these data were taken with different resolutions and signal-to-noise ratios, and since only equivalent widths are available to us for most of the comparison clusters, we are forced to use curve of growth techniques for abundance estimations.

The \ion{Li}{i} (6706.6\AA$+$6707.9\AA) doublet is blended with the \ion{Fe}{i} 6707.4{\AA} line only in lower resolution data. Where the Li doublet and Fe lines are resolved we use equivalent widths for the Li doublet alone if quoted. For Coma Ber stars the lines are clearly resolved in most of our WHY spectra and the equivalent width is determined by fitting a synthesised Li+Fe spectrum. Table 2 lists these EWs along with their errors determined from chi-squared fits. For lower resolution data and for our INT Coma Ber spectra we adopt a single deblending correction to the integrated Li+Fe EW. The empirical correction we use is that the EW of the Fe line (plus some even weaker CN features) is given by $20(B-V)-3$m\AA. In our INT spectra, the total Li$+$Fe EW is measured by integrating below a continuum, fitted to line-free regions around the Li doublet. These total EWs (and errors from the continuum fitting and integration) are listed in Table 2 along with the deblended Li doublet EWs. Note that the colour errors feed through to give a negligible additional error ($<$1m\AA) in the empirical deblending formula.

Of course, when dealing with clusters of differing metallicities we
might expect the deblending to vary. [Fe/H] for the Hyades is 1.5 times
that of Coma Ber and we would expect the weak blended Fe line to be 1.5
times as strong as well. We choose not to try and account for this
explicitly because (as we discuss further in Sect. 3.3) this effect
is small compared with the differences we see between the Hyades and
Coma Ber Li doublet EWs and besides, at the cool temperatures where the
deblending error might contribute most, the vast majority of the spectra are
taken at high resolution where the blending is not an issue.

Effective temperatures for all stars were derived from $B-V$ colours
using the calibration of Saxner \& Hammarb\"{a}ck (1985) for stars with
$B-V<0.64$ or B\"{o}hm-Vitense (1981) otherwise. The Saxner \&
Hammarb\"{a}ck relation includes a small metallicity dependent term and
we assume that a correction of the same additive form is appropriate
for use with the B\"{o}hm-Vitense relation. The effective temperatures
of the Coma Ber stars along with errors derived by propagating the
photometry errors through the colour-temperature relation are given in
Table 2. We note here that choice of effective temperature scale or the
metallicity dependence of the temperature scale cannot alter our
conclusions significantly. As we will see in Sect. 3.3 a change in
effective temperature changes the derived Li abundance such that a star
moves {\em along} the trend of Li versus effective temperature. So
while absolute Li abundances for individual stars might change, the
trends in abundance and inter-cluster comparisons do not.
Li abundances (quoted as $\log N(Li) = 12 + \log(Li/H)$) were estimated
using interpolation of the LTE curves of growth provided by Soderblom et al.
(1993a) - which use the appropriate gravity for the ZAMS stars considered here.
Small (of order 0.05-0.1 dex) NLTE corrections were made to these using
the code of Carlsson et al. (1994). The sign of these corrections varies with temperature: for hot stars it leads to a decrease in the Li abundance, while the reverse is true for cooler stars. For the Coma Ber stars the EWs,
adopted $T_{\rm eff}$ values and their associated errors are given in
Table 2. NLTE Li abundances are also given along with an error that is
derived purely by folding the EW and $T_{\rm eff}$ errors through the
curves of growth. Note that these errors do not reflect the true abundance
errors, which must also include contributions from uncertainties in the
temperature scale, atmospheres, convection treatment, microturbulence
etc. (see below). However they do represent the true levels of
uncertainty for comparing relative abundances between stars in the Coma
Ber and other clusters which have similar colours and Li abundances
determined using the same curves of growth.

Special mention is made of BD+28 2119.
This star has little photometry available and the Tycho $B-V$ value
had very large errors (0.876$\pm$0.125). We measured the
strength of several \ion{Fe}{i} lines in the 6708{\AA} region for all the
member stars with WHT spectra and plotted them against $B-V$. The
resulting graphs seem to indicate that the true $B-V$ for BD+28 2119 (assuming
it to be a typical cluster star) is closer to 0.79 $\pm$ 0.02 and this
is what we use to calculate the Li abundance.

In Sects. 4.1 and 4.2 we compare the Li abundances in Coma Ber (and
other clusters) with theoretical models of Li depletion. For this
purpose it is important to have an idea of the accuracy of the absolute
Li abundances, in addition to the precision of the relative Li
abundances. We expect larger, systematic errors in the absolute Li
abundances than the uncertainties listed in Table 2. This is because we
must include contributions from uncertain temperature scales,
microturbulence and the fact that different atmospheres and convection
models will yield slightly different abundances.  Changes in the
atmospheric assumptions do change the abundances in all the clusters
systematically with respect to the theoretical models. We have
experimented with this by performing a spectral synthesis for our best
WHT spectra using Kurucz {\sc atlas9} model atmospheres (Kurucz, \cite{kurucz93})
using different convection models (mixing length theory with and
without overshoot or full spectrum turbulence), different
microturbulence parameters and temperatures differing by $\pm 100$\,K.

We find that the abundances we determine are within 0.1 dex of the
abundances determined from curve of growth techniques and that alteration
of the atmospheric assumptions also causes Li abundances changes of 0.1
dex or less. We conclude that for the purposes of comparison with
theoretical models, systematic errors of about 0.1 dex in the Li abundance should be added to the relative errors listed in Table 2.

\section {RESULTS}

\subsection {Radial Velocities}

A summary of the heliocentric RV measurements found in this paper,
along with others from the literature, is given in Table 1. We take the
range of RVs for likely cluster members to be 2$\mathrm{\sigma}$ from $\pm$ 1
km s$^{-1}$ (O98), 
which is a typical cluster velocity dispersion (Rosvick et
al. \cite{rosvick92}). The cluster convergent point is sufficiently
distant that any change in RV with position will be negligible.
We suggest that those objects with radial velocities just outside our
range are likely to be widely-spaced spectroscopic binaries which are
members of the cluster. This is the case for Tr\,102, which 
has also been identified as an SB1 star by Bounatiro (\cite{bounatiro93}).

The combination of proper-motion, photometric, and radial-velocity
selection make it very unlikely that we have wrongly classified
non-members as cluster stars. For example, O98 estimate a
contamination level of less than 5 per cent among their brightest
stars (with Hipparcos proper motions), based on proper-motion
measurements alone. For fainter stars with less-accurate ACT proper
motions this probability is likely to be much larger (evidence
presented in Sect. 4.3 suggests it could be as high as 40 per
cent). However, if we assume field stars have RVs which are randomly
distributed between $\pm$30\,km\,s$^{-1}$ (Wielen 1977), then there is only a
$\sim10$ per cent chance of mistaking a non-member for a cluster
member using the RV classification alone, and this will shrink
drastically when we combine it with proper-motion and photometric selection.

\begin{table*}
\caption {Heliocentric radial velocity and projected equatorial
velocity measurements from this and previous papers}
\begin {tabular}{lllllllll}
\hline
Object 	& Other 	& RV from this 	& Other RVs &$v_{e}\,\sin i$	 & Status & Source \\
	& Names		& paper (km s$^{-1}$)		& and
references & (km s$^{-1}$)& & \\
\hline
\object{BD+16 2505} &		& $-$11.8 $\pm$ 2.0	&& $<15$	& NM		& INT98 \\
BD+21 2514 &	& $-$0.6 $\pm$ 2.0	&	& $<15$&M		& INT98 \\
\object{Tr 12} & \object{BD+26 2314}, \object{HD 105863}	& +1.2 $\pm$ 2.0	& +0.43(a)	& $<15$&M	& INT98 \\
\object{HD 114400} & \object{BD+34 2398} 	& $-$1.7 $\pm$ 2.0         &	& $<15$&M	& INT98	\\
\object{HIP 61205} &\object{BD+36 2278}, \object{Bou 50}	& $-$0.5 $\pm$ 2.0	& +1.1 (d)	& $<15$&M	& INT98	\\

\object{BD+21 2514} &	& 0.0 $\pm$ 1.0		& 	& $<6$& M		& WHT98 \\
\object{BD+25 2631} &		& +13.7 $\pm$ 1.0	&	& $<6$&NM		& WHT98 \\
\object{Ta 13}	& \object{BD+26 2342}, \object{Tr 213}	& +0.9 $\pm$
1.0	& $-$1.6 (d)	& $<6$&M	& WHT98\\
\object{Tr 102}	& \object{BD+27 2121}	& $-$3.1 $\pm$ 1.0	& $-$14,
var (a), $-$14 (c)	&$<6$& SB1 & WHT98	\\
\object{Ta 20} & \object{BD+27 2139}, \object{Tr 220}	& +0.5 $\pm$ 1.0	& +0.88 (a), +4.0 (d)	& $<6$&M	& WHT98	\\
\object{BD+28 2119} &		& +2.0 $\pm$ 1.0	&	&6& M
& WHT98	\\
\object{Tr 192} & \object{BD+28 2134}, \object{HD 109483} & $-$1.3 $\pm$ 1.0	& $-$2.4 (b)	&$<8$& M	& WHT98 \\
\object{Tr 141} 	& \object{BD+29 2290}	& +0.6 $\pm$ 1.0	&	& $<6$&M	& WHT98	\\
\object{BD+36 2312} &		& $-$55.6 $\pm$ 1.0	&	&8& NM	& WHT98	\\
\object{BD+38 2436} &		& $-$2.0 $\pm$ 1.0	&	&$<6$& M	& WHT98	\\
\object{Tyc 2534 1715 1}&       & $-$36.4$\pm$ 1.0        &       &$<6$& NM   & WHT98 \\
\object{HD 111878} 	& \object{BD+26 2402}		&$$-$2.4\pm1.0$& $-$7.5 (b)	& 10& M	& WHT99	\\
\object{Tr 86} 	& \object{BD+28 2109}, \object{HD 107611} &$$-$1.0\pm1.0$ 	& +2.4 (a), $-$0.7(c) & 20&M	& WHT99	\\
\object{Ta 3} 	& \object{Tr 203}			&$$-$15.9\pm1.0$ 	&	& $<6$&NM	& WHT99	\\
\object{Ta 19} 	& \object{Tr 219}	&$$-$14.9\pm1.0$  	&	& $<6$&NM	& WHT99	\\
\object{Ta 21} 	& \object{Tr 221}			&$+30.6\pm1.0$	&	& $<6$&SB1?	& WHT99	\\
\hline
\multicolumn{7}{l}{References: (a) Bounatiro 1993, (b) Knude 1989, (c)
Duflot et al. 1995, (d) Jeffries, 1999} \\
\multicolumn{7}{l}{Status: M -- Member, NM -- Non-member, SB1 -- Spectroscopic Binary}\\
\end {tabular}
\end {table*}

We stress that our assessment of membership is based only on
photometric, proper-motion and radial-velocity considerations. In
particular we do not consider the lithium abundance as a defining
characteristic. However, most of the stars which showed very discrepant
RVs (in this paper and J99) showed either very little or no lithium
compared with candidate members of similar $T_\mathrm{eff}$ (see Sect.
3.3). The exceptions here are the candidate SB1 system, Tr 102, and the
RV non-member, Ta 21, which {\em do} show Li abundances consistent
with those of other cluster members. Ta 21 {\em may} be a close binary
system in the Coma Ber cluster and warrants further RV investigation, but we
conservatively treat it as a non-member in the rest of this paper.  We
suspect that none of the remaining RV non-members are likely to have
discrepant velocities as a result of being members of close, 
tidally-locked binaries, because such systems usually have the same, or even
enhanced, Li relative to single cluster stars ({\em e.g.} those in the Hyades
-- Thorburn et al. \cite{thorburn93}).  We note that Argue \& Kenworthy
(\cite{argue69}) rejected Ta 3 and Ta 19 (selected as candidate
proper-motion members by Trumpler 1938) as cluster members on the basis of
their proper-motion measurements. We find that both these stars also
have RVs which are inconsistent with cluster membership, so the
photometric modulation observed by Marilli et al. (1997) must identify
them as moderately-youthful disk field stars.

\subsection{Projected equatorial velocities}

\begin{figure}
\resizebox{\hsize}{!}{\includegraphics{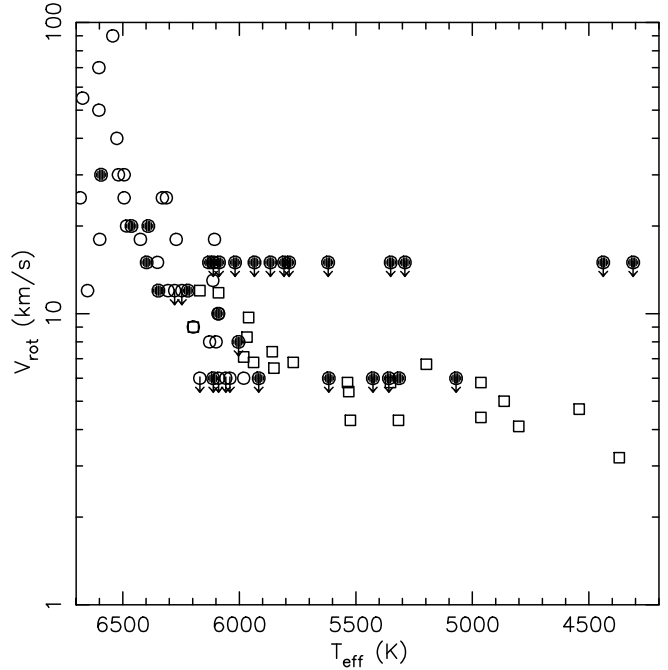}}
\caption{Rotational velocity versus $T_\mathrm{eff}$ for Coma Ber
(filled circles) and Hyades (open circles -- lower limits to true
equatorial rotational velocity (Kraft 1965); squares -- rotational
velocity calculated from rotation period and assumed radius (Radick et
al. 1987)).}
\label{rotplot}
\end{figure}

For the 20 stars in our sample, all have $v_{e}\sin i$ $\leq$ 20 km
s$^{-1}$, and at least 11 have $v_{e}\sin i$ $\leq$6 km
s$^{-1}$. Combining the Coma Ber cluster candidates from Table 1 with
those from J99 (where all candidates had $v_{e}\sin i$ $\leq$15
km\,s\,$^{-1}$) and the F stars with $v_{e}\sin i$ measurements in Kraft
(\cite{kraft65}), we can compare rotation rates in Coma Ber with those in the
Hyades. The Hyades measurements are either $v_{e}\sin i$ values from
Kraft (1965), or actual equatorial velocities derived from rotation
periods by Radick et al. (\cite{radick87}). The comparison is shown in
Fig.~\ref{rotplot} as a function of $T_\mathrm{eff}$ (calculated from
$B-V$ and [Fe/H] as described in Sect. 2.3),
and reveals no evidence for a difference in
rotation rates between Coma Ber and the Hyades. The equatorial
velocities in the Hyades show little scatter at a given $T_\mathrm{eff}$
and appear to decrease smoothly from about 20--30\,km\,s$^{-1}$ in mid-F
stars ($T_\mathrm{eff}=6500$K) to 10\,km\,s$^{-1}$ in late-F stars ($T_\mathrm{eff}=6100$K), and $\sim$5\,km\,s$^{-1}$ in late-G/early-K stars
($T_\mathrm{eff}=5000-5500$K). Recalling that the projected equatorial
velocities we observe have an extra (presumably random) $\sin i$ term
factored in, then the predominance of upper limits in our data is
quite consistent with the Hyades data.
 
\begin{table*}
\small
\caption{ Photometry, Blended (\ion{Li}{i} + \ion{Fe}{i}) and deblended \ion{Li}{i} 6708{\AA} EW, assumed T$_\mathrm{eff}$ values and Li abundances of stars in Coma Berenices. For the purposes of comparison with theoretical models, systematic errors of about 0.1 dex in the Li abundance should be added to the relative errors listed in Table 2. } 
\begin{tabular}{llllllll}
\\
\hline
Object & V & $B-V$ & Phot. & Blended  & Deblended & T$_\mathrm{eff}$ &
$\log N\mathrm{(Li)}$ \\
       &   &       & Ref.   &  EW(m{\AA}) & \ion{Li}{i} EW(m\AA)      & (K)       &       \\
\hline

BD+16 2505 & 9.62 & 0.658 $\pm$ 0.036 & tyc & 11 $\pm$ 3  & $\leq$ 7 &
5683 & $\leq$ 1.32 \\ 
\object{BD+21 2514}$^{(*)}$ & 10.12 & 0.747 $\pm$ 0.041 & tyc &  ... & 31.1 $\pm$ 1.7 & 5427 & 1.78 $^{+0.10}_{-0.10}$ \\
BD+21 2514$^{(\dag)}$ & 10.12 & 0.747 $\pm$ 0.041 & tyc & 56 $\pm$ 5 & 44.1 & 5427 & 1.93 $^{+0.11}_{-0.11}$ \\
BD+25 2631 & 10.48 & 0.848 $\pm$ 0.086 & tyc & ... & $\leq$ 3.5 & 5170 & $\leq$ 0.55 \\
HD 111878 & 8.87  & 0.546 $\pm$ 0.015 & hip & 85.0 $\pm$ 1.5
& 77.1 & 6091 & 2.73 $^{+0.04}_{-0.04}$ \\
Tr 12 & 9.54 & 0.624 $\pm$ 0.002 & hip & 82 $\pm$ 4 & 72.5 & 5808 & 2.49 $^{+0.03}_{-0.03}$ \\
Ta 13 & 10.51 & 0.773 $\pm$ 0.017 & d & & 25.0 $\pm$ 1.0 & 5358 & 1.62
$^{+0.04}_{-0.04}$ \\
Tr 102 & 9.36 & 0.594 $\pm$ 0.005 & d & 59.6 $\pm$ 1.4 & 49.7 & 5917 & 2.38 $^{+0.02}_{-0.02}$ \\ 
Ta 20 & 10.78 & 0.89 $\pm$ 0.01 & a & ... & $\leq$ 6.0 & 5070 & $\leq$ 0.69 \\
Tr 86      & 8.50  & 0.463 $\pm$ 0.005 & d & 44.0  $\pm$ 1.5&37.8  & 6391 & 2.60 $^{+0.02}_{-0.02}$ \\
BD+28 2119 & 10.50 & 0.79 $\pm$ 0.02 & e & ... & 28.8 $\pm$1.0 & 5316 & 1.64 $^{+0.05}_{-0.05}$ \\
Tr 192 & 9.07 & 0.57 $\pm$ 0.03 & tyc & ... & 69.8 $\pm$ 1.7 &6003 & 2.62 $^{+0.06}_{-0.06}$ \\
Tr 141 & 9.72 & 0.71 $\pm$ 0.06 & tyc & ... & 45.0 $\pm$ 1.2 &5616 & 2.04 $^{+0.12}_{-0.12}$ \\
HD 114400 & 9.59 & 0.608 $\pm$ 0.03 & tyc & 66 $\pm$ 4 &  57.2 & 5866
& 2.41 $^{+0.09}_{-0.09}$ \\
HIP 61205 & 9.71 & 0.63 $\pm$ 0.01 & a & 77 $\pm$ 4 & 67.4 & 5786 & 2.43 $^{+0.05}_{-0.04}$ \\
BD+36 2312 & 10.47 & 0.86 $\pm$ 0.01 & b & ... & $\leq$ 3 & 5142 & $\leq$ 0.45 \\
BD+38 2436 & 9.10 & 0.54 $\pm$ 0.01 & c & ... & 80.7 $\pm$2.0 & 6112 & 2.77 $^{+0.03}_{-0.03}$ \\
Ta 3       & 10.77 & 0.85 $\pm$ 0.01 & a & ... &  $\leq$ 5.8  & 5167 &  $\leq$ 0.77 \\
Ta 19      & 11.18 & 0.82 $\pm$ 0.01 & a  & ... & $\leq$ 5.6  & 5240 & $\leq$ 0.83  \\
Ta 21      & 10.36 & 0.81 $\pm$ 0.01 & a & ... &  15 $\pm$ 2.0  & 5265  &  1.30$ ^{+0.02}_{-0.02}$ \\
Tyc 2534 1715 1& 10.42& 0.797 $\pm$ 0.078 & tyc & ... & $\leq4.4$   &5298&$\leq0.77$    \\
\hline
\multicolumn{8}{l}{References: (a) Mermilliod (1976); (b) Haagkvist \&
Oja (1973) ; (c) Oja (1985); (d) Johnson \& Knuckles (1955) } \\
\multicolumn{8}{l}{(e) see Sect. 2.3; hip -
Hipparcos catalogue; tyc - Tycho catalogue; $^{*}$ WHT spectrum; $^{\dag}$ INT spectrum}\\

\end {tabular}

\normalsize
\end{table*}

\subsection {Lithium}

The photometry, Li + Fe EWs, deblended EWs, assumed T$_\mathrm{eff}$ values, 
and calculated Li abundance for each star are given in Table 2. Errors
in the Li abundances take into account random errors in both the
photometry and Li EWs. These are internal errors and additional external errors of $\pm$0.1 dex are appropriate when comparing with Li abundances determined using different techniques. Three of our measurements can be compared with
those of J99 (based on spectra obtained with the INT in an identical
configuration to that used here):
\begin {itemize}
\item HIP 61205 (Bou 50). Our measurement for this
star was also taken on the INT, and is in excellent agreement with the
value from J99. This suggests that there is no systematic difference
between our reduction and that of J99, at least for the INT spectra
\item Ta 13. The deblended Li EW estimated in J99 is
28$\pm3$m\AA, in excellent agreement with the measured WHT Li EW 
in this paper. This gives some support for the accuracy of the
empirical deblending performed both in this paper and in J99.
However, we also note that in the case of BD+21 2514, for
which we have obtained both INT and WHT spectra, there is a
discrepancy of (14.9$\pm$5.3)m\AA\ between the empirically-deblended
INT Li EW and the measured WHT Li EW. The question of the detailed
accuracy of the empirical-deblending process is still open and we have
much more faith in the measured Li EWs from the higher resolution WHT
spectra. But, as discussed in Sect. 3.2 of J99, errors in the
empirical deblending will not alter the conclusions in that paper or this one.
\item Ta 20. The higher resolution WHT spectra have
allowed us to obtain a new upper limit value for
the \ion{Li}{i} 6708{\AA} EW for this star, which is smaller than the
previous value by 6m{\AA}.
\end {itemize}

\begin{figure*}
\resizebox{\hsize}{!}{\includegraphics{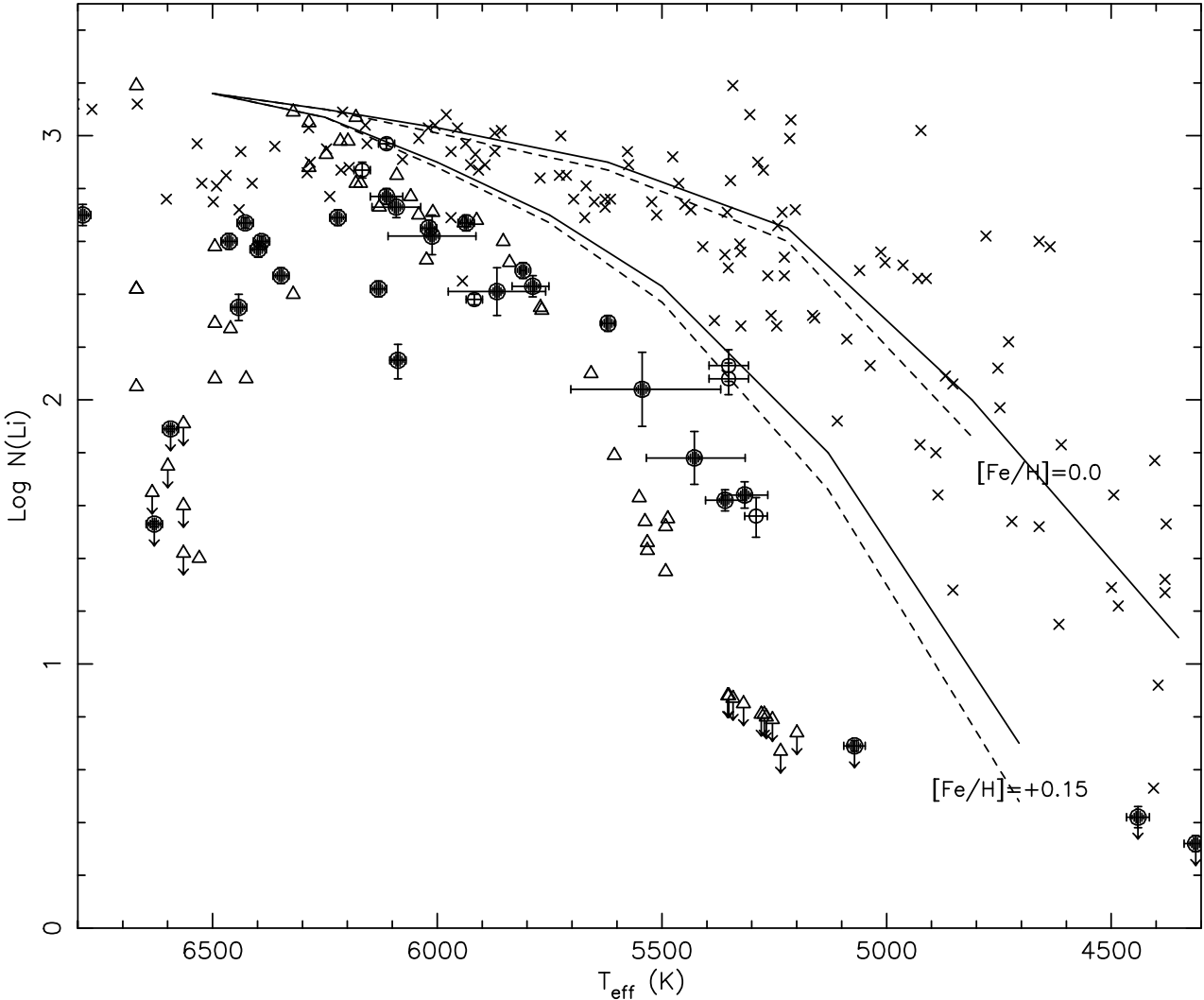}}
\caption{NLTE Li abundances versus $T_\mathrm{eff}$ for Coma Ber (filled
circles -- single stars, open circles -- spectroscopic binaries),
Hyades (triangles) and Pleiades (crosses) open clusters. The two solid
lines are standard model isochrones of Li depletion (from Pinsonneault
(1997)) for and age of 100Myr and [Fe/H] of 0.0 (upper) and +0.15
(lower). The dashed lines are as above but for an age of 700Myr.}
\label{ali}
\end{figure*}

A plot of Lithium abundance versus $T_\mathrm{eff}$ for the Coma Berenices
cluster members is shown in Fig.~\ref{ali}, where we have added the points from
J99 (favouring measurements done at higher resolution on the WHT), and
supplemented these with additional F stars observed by Boesgaard
(1987).  Open symbols represent those candidate members we assume to be
in binary systems. Comparison is made with similar data for
stars in the Hyades and Pleiades clusters. Li EWs and photometry were
obtained from Boesgaard \& Budge (1988), Soderblom et al. (1993b),
Thorburn et al. (1993), and Soderblom et al. (1995). In these
comparison samples we have included only those stars with no evidence
for close binarity. In all cases, where necessary, we first deblended
the Li line and then obtained Li abundances using the same temperature 
scale and curves of growth as for the Coma Ber data.

Fig. ~\ref{ali} shows that Coma Ber stars with spectral types earlier than that of the Sun
(5800\,K$<T_\mathrm{eff}<$6700\,K) are clearly Li-depleted when
compared with the Pleiades.  This depletion is most dramatic at around
6600\,K, the ``Boesgaard gap'', which is more clearly seen in the
Hyades data. The abundances in the Coma Ber F and early-G stars appear
similar to those in the Hyades, except for a suspicion that some Coma
Ber stars with temperatures around 6100--6400K may exhibit 0.3--0.8 dex
{\em more} depletion than their Hyades counterparts, although this comparison
is hampered by small numbers. However, there are three Coma Ber
stars which have clearly depleted their lithium abundances below the
Hyades level. The Hyades stars show little
evidence for a lithium abundance dispersion at a given temperature in
this region, but the presence of some single and probable-binary stars with Li
abundances similar to those of Hyades stars indicates a large spread in
these abundance values among the late-F stars of Coma Ber.

The three late-F low-Li Coma Ber stars are Tr\,58, Tr\,76 and
Tr\,92. Tr\,58 has consistently measured abundances given in both J99 and
Boesgaard (1987). Tr\,76 has consistently low Li measured by
J99, Boesgaard (1987) and Soderblom et al. (\cite{s90}). Tr\,92 has
consistent measurements in both Boesgaard (1987) and Soderblom et
al. (1990).  Either there is a genuine large spread in Li depletion for
late-F stars in Coma Ber (which is not seen in the Hyades) or these
three stars coincidentally show proper motions, photometry and radial
velocities consistent with cluster membership -- which we think very unlikely.

For stars with $T_\mathrm{eff}<5800$\,K, the addition of several new Coma
Ber members adds weight to the tentative finding by J99, that the
abundances of Coma
Ber stars are close to, but have suffered less Li depletion than, stars
of similar $T_\mathrm{eff}$ in the Hyades. This discrepancy widens
at lower $T_\mathrm{eff}$.  The reader might imagine that uncertainties in
the metallicities of the two clusters, combined with the metallicity
dependence of the $T_\mathrm{eff}$-$B-V$ relation, could partially account
for the difference. At $B-V=$0.8 the Hyades stars are around 100\,K hotter
than those in Coma Ber for the metallicities we have assumed. However,
if the Coma Ber temperatures {\em were} higher, the derived Li
abundances would also increase. The net effect would be to move the
points almost parallel to the trend observed in the G and K stars of
Coma Ber and the Hyades, thus making little difference to the
comparison.  A further objection could be that the empirical deblending
formula used for the INT data must surely be metallicity dependent. Of
course Coma Ber has the lower metallicity, so the contribution to the
blended Li+Fe line should be {\em smaller} than for the Hyades, by
a factor of around 1.5 for this weak line. This would have a small effect
in the sense of widening the discrepancy between Coma Ber and the
Hyades. Besides, we now have higher resolution data for most of the 
cool stars, where the line blend is resolved (see Fig.~\ref{spec}). 
Therefore, we believe that this difference in Li depletion is real and well 
established, since it is now supported by several more objects.

\section {DISCUSSION}

For the purposes of discussion we divide our sample into two temperature ranges. We stress that the Li abundances listed in Table 2 (and for stars in other clusters) have been determined using consistent methods. Thus intercomparison of the cluster samples need only consider internal errors due to photometry and EW uncertainties. However, where we make comparison with theoretical models it must be accepted that there is an additional $\sim\pm0.1$dex error in the absolute Li abundances. In addition we assume that all clusters were born with the same initial Li abundance, but variations of $\pm$0.1 dex are possible.

\subsection {F-stars (5800 $\leq{T_\mathrm{eff}}\leq$ 6800\,K)}

What expectations do we have for the Li depletion patterns of F stars
in clusters of similar age but differing metallicity? Our observations
suggest that there is little difference in the position of
the centre of the Boesgaard gap or its cool wing. This is in agreement
with the work done by Balachandran (1995), who showed that there was
essentially no difference between the Hyades and Praesepe F-star
Li-depletion patterns, despite a small metallicity difference between
the clusters. We do, however, see some evidence for additional Li
depletion, relative to the Hyades (and Praesepe), among some (but not
all) Coma Ber stars with temperatures in the range 6100--6300\,K.

Standard models predict that we should see almost no difference in the
Li-$T_\mathrm{eff}$ trends of the Hyades, Praesepe or Coma Ber F stars,
since their CZs are thin enough that convective PMS Li destruction
amounts to only $\sim$0.1 dex. Still smaller discrepancies are seen 
between models with metallicity differences of 0.1--0.2\,dex
(Pinsonneault 1997). However, we see significant Li depletion among
the F stars (6000--6700\,K) in both the Hyades and Coma Ber ranging
from 0.3 dex to $>$1.5 dex in the Boesgaard gap (for an assumed initial $\log N\mathrm{(Li)}$ of 3.2).
This depletion, which is extra to the standard model predictions,
demands non-standard effects such as microscopic diffusion or slow
(non-convective) mixing caused by rotation or gravity waves. 

Boesgaard (1991) presented an interesting argument, suggesting that for
stars of a given age which are in the temperature range 5950--6350\,K, 
greater Li depletion would be seen in lower metallicity objects
because the thinner CZs should make microscopic diffusion more
efficient. This would lead us to expect significantly greater
depletion in the Coma Ber stars compared to those of the Hyades. This
is apparently what we are seeing and a $\log N\mathrm{(Li)}$ value of
2.7 in the late-F stars of Coma Ber would be in reasonable agreement
with the empirical relationship between Li abundance, age, and
metallicity put forward by Boesgaard. Whilst we confirm this empirical
trend, we do not  believe that Boesgaard's explanation is correct. No
account is taken of the fact that for late-type stars of a given
temperature, a lower metallicity star has a lower mass and a
relatively thicker CZ. These two effects almost cancel out for late-F
stars, so that stars of a given $T_\mathrm{eff}$ have similar CZ
thicknesses. It is thus very unclear how a small change in metallicity
affects Li depletion in the context of the microscopic diffusion model
and it would be interesting to see some numerical predictions.

The work of Deliyannis et al. (\cite{deliyannis98}), argues strongly
against the effectiveness of diffusion in late-F stars. Instead, their
observations of simultaneous Li and Be depletion in field F stars
favour rotationally-driven mixing occuring between the CZ base and
regions where the temperature is high enough to burn Li.  Models
incorporating rotational mixing are described by Pinsonneault et
al. (1999). These show that in the range 6000--6500\,K, the predicted
depletion by the age of the Hyades is modestly metallicity dependent,
with about 0.1--0.2 dex {\em less} depletion in Praesepe than in the
Hyades. Intuitively this arises because at a given $T_\mathrm{eff}$, the
CZ base is cooler in lower metallicity stars, so material has to be
mixed further to deplete Li.  Extrapolating, we might expect even less
depletion in Coma Ber stars at the same temperature, so the observed Li
abundances do not seem to agree with the predictions for rotational
mixing. However, more important than metallicity in these models is
the rotational history of the stars themselves. Stars which begin life
rotating more rapidly are able to deplete their Li faster. Hence a
wide spread in initial rotation rates can lead to a wide spread in Li
depletion at later times. We could hypothesize that the Coma Ber stars
began life rotating slightly faster on average than Hyades stars, and
with a larger spread in initial angular momentum. 

The current rotation rate data in these clusters (shown in
Fig.~\ref{rotplot}) neither support nor contradict this hypothesis.
The suggestion is difficult to test because: (a) We only have
$v_{e}\sin i$ measurements in Coma Ber, many of which are upper
limits. However, the three anomalously low Li F stars in Coma Ber do
not have anomalously high $v_{e}\sin i$. (b) Magnetic braking causes
rotation rates to converge among stars with outer convection zones, so
a lack of a rotation-rate spread now, does not mean that one never
existed. Certainly, spreads in Li abundance are clear among the F and G
stars of older clusters such as M67 (Jones et al. 1999), and one
possible explanation which has been put forward is differing
Li depletion rates caused by different rotational histories.

\subsection{G and K stars ($T_\mathrm{eff}\leq$\,5800\,K)}

Our observations of G and K stars in the Coma Ber cluster cannot be
adequately explained by either standard or non-standard
models. Standard models predict that almost all depletion in these
stars should happen on the PMS, with no further Li destruction once
stars reach the ZAMS (Pinsonneault 1997 and Fig.~\ref{ali}). The amount of PMS
depletion will be highly dependent on the metallicity of the star:
metal-rich stars will have deeper CZs and therefore deplete Li more
efficiently than lower-metallicity stars of the same temperature. 
Recalling the metallicities of Coma Ber ($-$0.052$\pm$0.026), the
Pleiades ($-$0.034$\pm$0.024) and the Hyades (+0.127$\pm$0.022), this would mean that the Coma Ber lithium abundances should be close to those of
similar stars in the Pleiades, due to their comparable metallicities. 
To illustrate the dependence of PMS depletion on metallicity,
Fig.~\ref{ali} shows two sets of Li depletion isochrones with a
metallicity difference of 0.15 dex, approximately the difference 
between the Hyades and the Pleiades.

In fact, the Coma Ber abundances are closer to those of
Hyades stars which have a similar age, but a significantly higher
metallicity (by $0.179\pm0.034$ dex). The
difference between the Coma Ber and Pleiades trends can only be
explained by MS mixing, ruling out the standard model. The amount of
mixing required is made uncertain by the errors in the cluster
metallicities, but must give rise to around 0.3 dex of Li depletion at
6000\,K, rising to 0.8--1.2 dex at 5400\,K. 

The discrepancy between Coma Ber and the Hyades, which was tentatively
identified by J99 and which is now confirmed by our larger sample,
seems far too small to be explained by standard PMS depletion. Any MS
depletion of lithium should reduce the amount of Li in {\em both}
clusters, without decreasing the abundance gap between them, unless
initial rotation rates in Coma Ber were considerably larger than in
the Hyades. The evidence shown in Fig.~\ref{rotplot} does not support
higher rotation rates in Coma Ber, although some convergence of
rotation rates may have taken place over several hundred million years.
If anything, we might expect MS mixing to amplify the difference between
the cluster Li-$T_\mathrm{eff}$ trends, since the more metal-rich
Hyades stars should destroy their Li more efficiently than those of
Coma Ber. The Hyades is also $\sim$ 100--200 Myr older than Coma Ber
allowing more MS mixing to take place. This suggests that
Coma Ber and the Hyades started life  on the ZAMS with Li abundances
which were similar (at least for stars in the temperature range
5400$<T_\mathrm{eff}<$6000\,K), and comparable with the level of Li seen in the Pleiades today. Again, this flatly
contradicts the predictions of standard models but a lack of metallicity
dependence for PMS Li depletion is also supported by observations of
the young metal-rich cluster Blanco 1 (Jeffries \& James 1999) and the young metal-poor
cluster NGC 2516 (Jeffries et al. 1998). Both of these clusters show
ZAMS Li depletion patterns and rotation rate distributions similar to that of the Pleiades. 

There are two possibilities remaining that could rescue the standard
model of PMS Li depletion. The first is that the metallicities of Coma
Ber, the Hyades and the Pleiades are incorrect. Whilst we believe the
absolute [Fe/H] values we have used probably have errors larger than those
stated, the quoted errors probably do reflect the uncertainties in the
{\em relative} abundances of the three clusters, because they were
analysed using similar spectra, atmospheric models and temperature
scales.  
The second possibility is that these clusters have non-solar
abundance ratios. The metallicities we have used refer only to the Iron
abundance, whereas CZ structure, and hence Li depletion, is
also significantly affected by the abundances of Silicon, Neon,
Magnesium and Oxygen (Swenson et al. \cite{swenson94}).  A detailed and
consistent abundance analysis for all these elements in the relevant
clusters would be invaluable. However, we consider this unlikely to be a
saviour for the standard model simply because of the growing weight of
evidence that cluster Li depletion patterns among G and K stars are
ordered according to their age rather than their metallicity (Jeffries
\cite{jeffries00}).

\subsection {The cluster luminosity function and mass segregation} 

\begin{figure}
 \resizebox{\hsize}{!}{\includegraphics{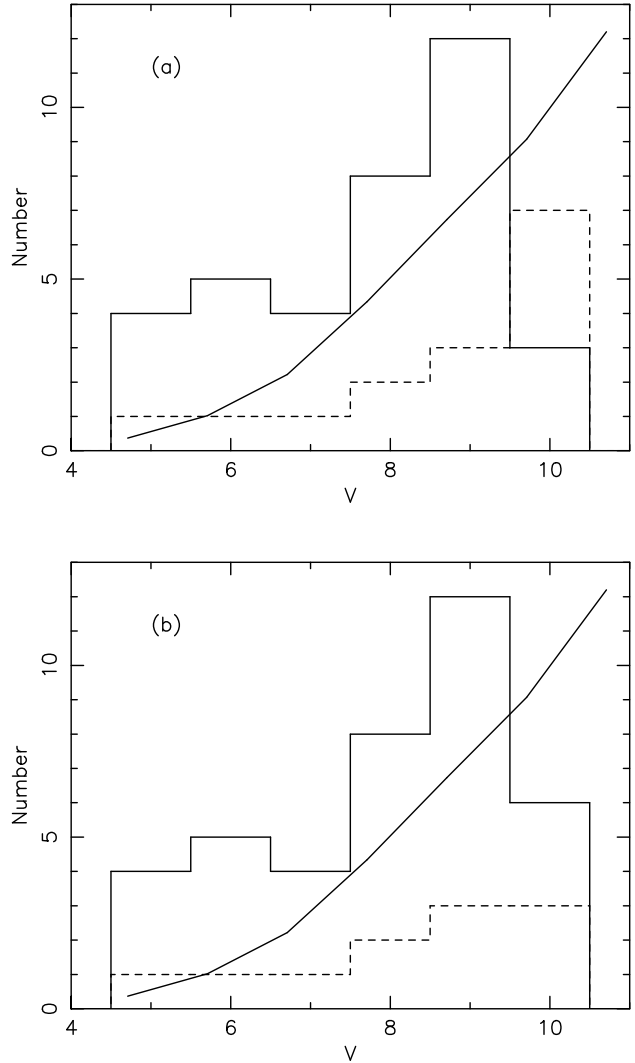}}
 \caption{Luminosity functions for the central parts of the cluster
(solid histogram), the moving group (dashed histogram) and a
normalised canonical field luminosity function from Miller \& Scalo
(1979 -- solid line). (a) Comparison based on the original sample and
classification presented by O98. (b) Comparison
after the samples have been modified to include new members and
exclude new non-members found in this paper.}
 \label{lf}
\end{figure}

In Sect. 5 of their paper, O98 examine the
spatial and kinematic distributions of proper-motion selected Coma Ber
candidates. They concluded that the cluster core is surrounded by a
moving group of extra-tidal stars which share similar space motions
but which are not gravitationally bound to the system as a whole. 
By separating the cluster candidates into two subsets (those objects
within 5 degrees of the cluster centre, and those more than 5 degrees
from it), it was shown that whilst the luminosity function falls
sharply at  $V=9$ in the centre, the moving group still appears to
have a rising luminosity function at $V=10$ (see Fig.~\ref{lf}). The
inferences from this are that the drop in the cluster-core luminosity
function was real and not caused by incompleteness, and that the origin of
the moving group is low-mass stars which have suffered dynamical
segregation and evaporation from the cluster core. We could also surmise that
even more low-mass members might be found among the moving group at
magnitudes below the current limits of proper-motion surveys.
Plots of both luminosity functions are given in Fig.~\ref{lf}a.

These conclusions hinge critically on the number of faint moving-group
proper-motion candidates which are {\em bona fide} cluster
members. There are only seven faint ($V>$9.5) moving-group objects,
(those which are more than 5 degrees from the cluster centre), and
they have all had their membership credentials tested by radial
velocity and Li abundance measurements in this paper. Four were found
to be non-members (BD+16 2505, BD+25 2631, BD+36 2312 and TYC 2534 1715
1), while the three others (BD+21 2514, HD 114400 and
HIP 61205) were confirmed as cluster members. There are also three
faint cluster-centre objects (BD+28 2119, Tr 12 and Tr 141) which we
have confirmed as members, J99 identified a further three stars with
9.5$<V<$10.5 which also appear to be cluster members, yet which are not
listed as such by O98: \object{Tr 120}, \object{Tr 132} and \object{Tr
150}. All of these would be classified as central-cluster members. 
This confirms that there is some incompleteness in the O98 sample and
weakens the evidence for a sharp drop in the luminosity function at $V=9$.

We have performed a formal Kolmogorov-Smirnov double-sided test on
the cumulative luminosity functions of the central cluster and moving
group from O98, using their membership
classifications. We find that the distributions are different at the 92
per cent confidence level in agreement with their conclusion.
However, if we remove the 4 stars which we classified as non-members
from the faint moving-group sample, this confidence level reduces to
45 per cent. Inserting the three 'new' members of the central cluster
reduces this confidence level further, to only 11 per cent, i.e. we
can conclude that the two luminosity functions are indistinguishable.
One-sided Kolmogorov-Smirnov tests of these modified cumulative
luminosity functions against the canonical field luminosity function
given in Table 1 of Miller \& Scalo (1979, assuming a cluster distance
modulus of 4.7), yield probabilities of 0.1 and 80 per cent
respectively that the central-cluster and moving-group luminosity
functions could be similar to those of field stars. Of course, these
probabilities depend critically on whether the sample is more
incomplete at fainter magnitudes. It would require adding a further
5--6 faint stars to the central-cluster sample to make its luminosity
function indistinguishable from that of the field. The implications of
this are, depending on the level of incompleteness, that the
luminosity function {\em probably} flattens, or even drops, for $V>9$ in
the central part of the cluster. This is presumably as a result of dynamical
evolution. However, there is no clear evidence that the evaporated or
ejected stars form the moving group around the cluster, or that the
luminosity function continues to rise among the faint moving-group
members: its luminosity function could be consistent with either the 
field or the central part of the cluster. Therefore, it
remains highly uncertain whether a significant number of moving-group
members could be found at fainter magnitudes and lower masses.

\section {CONCLUSIONS}

In this paper we have obtained radial velocities, projected equatorial
velocities and lithium abundances for a sample of Coma Berenices
open-cluster candidates, which were selected on the basis of their
proper motions. The radial velocities were used to refine the
candidate list further. Of the 20 objects observed, 6 were found to be
non-members, 1 is a spectroscopic binary, 1 is a probable
spectroscopic binary, while the remaining 12 have radial velocities
consistent with cluster membership. 

By combining our data with the work of O98, Boesgaard (1987) and Kraft
(1965), we have obtained a near-complete ($>$60 per cent)
membership census of F, G and early-K stars in the Coma Ber
cluster. We have examined Li depletion in these stars, compared them with
similar objects from the Hyades and Pleiades clusters and arrived at
the following conclusions: 
\begin {itemize}

\item Li depletion has occurred in the F stars of Coma Ber. The Li
levels in these stars are close to, or slightly lower than, those of
comparable stars in the Hyades, a cluster with a similar age but
higher metallicity. Several late-F stars in Coma Ber have suffered
0.2--0.5 dex more depletion than similar stars in the Hyades. 
It is unlikely that these stars are non-members and we have no
convincing explanation for this observation at present. We speculate
that some Coma Ber F stars had a significantly different rotational
history to those in the Hyades, although the rotation data in this
paper and the literature do not show any peculiarities to support this. 

\item We have confirmed the tentative results presented by Jeffries (1999)
with our larger sample of G and K objects in the Coma Ber cluster. We
find that these stars have undergone less Li depletion than their
counterparts in the more metal-rich Hyades, but significantly more
depletion than younger Pleiades stars of similar metallicity. These
observations contradict the standard Li-depletion model, consisting
mainly of PMS convective mixing, since this would predict a
strong metallicity dependence in Li-depletion patterns which we do not
see.

\item The difference between the Pleiades and Coma Ber Li depletion at
a given temperature could be attributed to non-standard mixing
mechanisms, such as rotation-induced
turbulence. If this is the case, such mixing must be
responsible for Li depletion of around 0.3 dex (at 6000\,K), rising to
0.8--1.2 dex in cooler objects (5400\,K).

\item The similarity between the Hyades and Coma Ber Li-depletion
patterns among G stars is difficult to explain with non-standard
mixing, unless both clusters joined the ZAMS with Li depletion
patterns similar to that of the Pleiades. Again, this contradicts our
notions of how Li depletion and mixing work during PMS evolution.
\end {itemize}

Finally, we have used our data to re-evaluate some of the conclusions
on the spatial distribution and luminosity function of proper-motion
cluster candidates which were reached in earlier work. In particular we have
found that, among the fainter stars, earlier membership samples were
not complete and included a significant number of non-members. 
Taking this into account we find that the luminosity functions of
stars in both the central cluster and in the proposed extra-tidal
moving group of star are indistinguishable for 0$<M_\mathrm{v}<$6. This
questions whether the moving-group is preferentially populated by
ejected or evaporated low-mass stars from the central cluster. However, there
is still evidence from the central-cluster luminosity function
that preferential evaporation of lower mass stars has indeed occurred.

\section {Acknowledgements}
The William Herschel and Isaac Newton Telescopes are operated on the
island of La Palma by the Isaac Newton Group in the Spanish Observatorio
del Roque de los Muchachos of the Instituto de Astrofisica de
Canarias.  The authors acknowledge the travel and subsistence support
of the UK Particle Physics and Astronomy Research Council (PPARC).  AF
and JRB were funded by PPARC postgraduate studentships. Computational work was
performed on the Keele and St. Andrews nodes of the PPARC funded Starlink
network. This research has made use of NASA's Astrophysics Data System
Abstract Service and CDS' SIMBAD Astronomical Database.

\end {document}